\documentclass[showpacs,amsmath,amssymb,twocolumn,floatfix,prl]{revtex4}
\usepackage[dvips]{graphicx}
\input{epsf}

\begin{document}

\title{Mesoscopic ``Rydberg'' atom in a microwave field}


\author{Toma\v z Prosen$^{(a)}$ and 
Dima L.~Shepelyansky$^{(b)}$}
\affiliation{$^{(a)}$ Physics Department, Faculty of Mathematics and Physics,
University of Ljubljana, Ljubljana, Slovenia \\
$^{(b)}$Laboratoire de Physique Th\'eorique, 
UMR 5152 du CNRS, Universit\'e P. Sabatier, 31062 Toulouse Cedex 4, France
}

\date{November 5, 2004}

\begin{abstract}
We establish analogy between a microwave ionization of Rydberg
atoms and a charge transport through a  chaotic quantum dot
induced by a monochromatic field in a regime with a potential
barrier between dot contacts.  
We show that the quantum coherence leads to dynamical localization
of electron excitation  in number of photons absorbed inside the dot.
The theory developed  determines the dependence of  localization length
on dot and microwave parameters  showing that the microwave power
can switch the dot between metallic and insulating regimes.
\end{abstract}


\maketitle
The dynamical localization of quantum chaos
is a generic physical phenomenon
induced by  quantum coherence which leads to
suppression of diffusive wave packet spreading
originated from dynamical chaos in the classical limit
(see e.g. \cite{dyloc} and Refs. therein).
It has been first seen in numerical simulations
of the kicked rotator model \cite{casati79}
and, later, of a more realistic system
of excited hydrogen atom in a microwave field
\cite{dls83}. The first observation  of this
phenomenon was achieved in microwave ionization 
experiments of hydrogen and Rydberg atoms 
\cite{koch,bayfield,walther,gallagher} while 
more recent experimental progress
allowed to realize the original kicked rotator
model with cold atoms in laser fields
and detect with them the 
dynamical localization of chaos \cite{raizen}. 

For a hydrogen atom in a microwave field
the diffusive excitation in energy
appears only above certain field threshold
where the integrability of unperturbed
motion is destroyed and the   
classical dynamics becomes chaotic \cite{delone}.
However, in certain systems the internal
dynamics can be fully chaotic on an energy surface 
so that classically the diffusive excitation
starts for arbitrary small fields.
As an example of such systems we may
quote complex molecules \cite{cederbaum}, Rydberg atoms
in a magnetic field \cite{wintgen}
or chaotic Sinai billiards \cite{sinai}.
In such systems the quantum eigenstates are chaotic and
their level spacing statistics is described by the 
random matrix theory \cite{bohigas,berry,cederbaum,wintgen}.
The excitation of such quantum systems by a monochromatic
field represents a diffusive process of one-photon absorption/emission 
transitions with a rate $\Gamma$ which is given  by the Fermi golden rule. 
The quantum interference effects lead to the
dynamical localization of this diffusion with
the localization length $l_\phi$ given by \cite{dls1987}:  
\begin{equation}
\label{Eq1}
l_\phi = 2 \pi \hbar \Gamma \rho_c \; .
\end{equation}
Here $l_\phi$ is measured in the number of photons of
monochromatic field with frequency $\omega$,
$\rho_c$ is the level density of unperturbed system
and it is assumed that $\hbar \omega \rho_c > 1$
and $l_\phi > 1$.
The extensive numerical simulations  of a
microwave ionization of internally chaotic Rydberg 
atoms (chaos is produced by static magnetic or  electric field)
\cite{dec3} showed that the relation (\ref{Eq1})
works even in extreme regimes
when up to a thousand of photons is needed
to ionize one atom.

The result (\ref{Eq1}) is rather general and can be used
not only for atoms but also for chaotic billiards
in a microwave field. In fact, during the last decade
the conductance properties of such billiards
realized with 2D electron gas quantum dots of micron size 
have been studied in great detail both experimentally
and theoretically (see reviews \cite{beenakker,alhassid}
and Refs. therein). In such experiments
typical parameters
correspond to a dot size $a \sim 1 \mu m$ and
electron density $n_e \sim 4 \cdot 10^{11} cm^{-2}$.
With the spin degeneracy the level number 
at the Fermi energy is $n_F=n_e {\cal A}/2 \approx 2000$
where the dot area is ${\cal A} \sim a^2$. 
These values of $a$ and $n_F$ correspond to those
of  a Rydberg atom with the principal number
$n \sim 100$. Indeed, the atom size is
$a \sim n^2 a_B \sim 0.5 \mu m$
and the level number, inside a set with fixed
magnetic quantum  number $m_z \sim 1$, is
$n_F \approx n^2/2 \sim 5000$ (see e.g.\cite{dec3}). Here $a_B$
is the Bohr radius and $m_z$ is preserved
for a linearly polarized field.
Thus, a quantum dot with the above parameters
can be viewed as a mesoscopic ``Rydberg'' atom.
A tunneling barrier of large hight $U$ between the dot and
the outcome lead (see Fig.~1) creates an effective 
ionization potential similar to the one of Rydberg atom.
A small barrier $U_{in}$ between the dot and  income lead 
can be used to control electron confinement inside the dot.
The microwave field  will switch  
the dot conductance  from  insulating 
($ \hbar \omega l_\phi \ll U$) to  metallic
($\hbar \omega l_\phi > U$) regime.
 
To understand in a better way the requirements for
microwave parameters we note that 
for the 2D electron  gas the Fermi momentum $p_F$ and energy $E_F$
are given by $p_F^2 = 2 \pi n_e \hbar^2$ and 
$E_F =p_F^2/2m$ with $m=0.067m_e$ for $nGaAs$. The average level spacing
in the dot is $\Delta =1/\rho_c=E_F/n_F =2\pi \hbar^2/m{\cal A}$. 
For the typical
values of $a$ and $n_e$ used above we obtain $E_F \sim 100 K$,
$\Delta \sim 0.05 K$ with the Fermi velocity 
$v_F = p_F/m \sim 2 \cdot 10^7 cm/s$.
Hence, the collision frequency is
$\nu_c/2\pi =  v_F/2a \sim 100 GHz$ that is much larger than 
the frequency corresponding to one level spacing
$\Delta/(2\pi \hbar) \sim 1 GHz$. 
Thus for a microwave field with $\omega/2\pi < 100 GHz$
{\it ac-}driving is in the classical adiabatic regime
and according to the adiabatic theorem
the energy excitation is exponentially small
for dot billiards with integrable dynamics.
However, for dots with chaotic dynamics
a microwave field with $\hbar \omega > \Delta$
leads to diffusive excitation in energy $E$ with the rate
$D_E = (\delta E)^2/\delta t \approx \hbar^2 \omega^2 \Gamma$.
In analogy with a chaotic atom \cite{dec3} in a
linearly $x$-polarized field we have 
$\partial E/\partial t =  \epsilon \omega x \cos \omega t$ and
$D_E \sim (\epsilon a \omega)^2/\nu_c$
where $\epsilon$ is the field strength multiplied 
by the electron charge.
This gives 
\begin{equation}
\label{Eq2}
l_\phi \approx 2 \pi \chi \epsilon^2 a^2/(\hbar \nu_c \Delta) 
\approx 16 \chi \epsilon^2 ({\cal A}/{\cal A}_0)^{5/2}(n_{e0}/n_e)^{1/2} 
\end{equation}
where $\chi$ is a numerical constant
and in the right part ${\cal A}_0 = 1\mu m^2$, 
$n_{e0}=4 \cdot 10^{11} cm^{-2}$ and $\epsilon$ is measured in $V/cm$. 
It is also convenient to write (\ref{Eq2}) in the form
$l_\phi \approx \chi (\epsilon a/E_F)^2 n_F^{3/2}$
which shows that $l_{\phi}$ may be large even when 
$\epsilon a \ll E_F$.
\begin{figure}
\includegraphics[width=\columnwidth,angle=0]{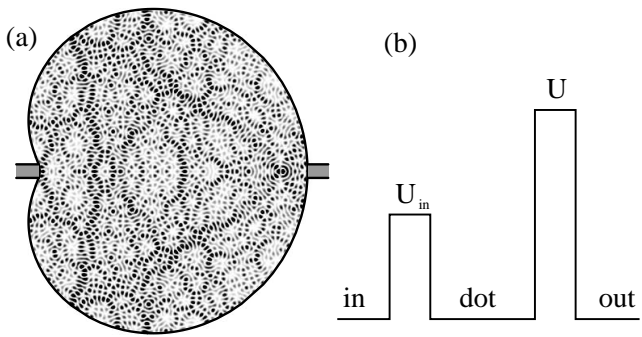}
\vglue -0.3cm
\caption{\label{fig1} (a) Dot billiard at $\lambda=3/8$
with the probability density
of the eigenstate at the Fermi level $n_F=2001$
(see text); density is proportional to grayness. Dashed stripes
show schematically income (left) and outcome (right) leads. 
(b) Sketch of tunneling
barriers $U_{in}$ and $U$ between income lead,  dot and 
outcome lead . 
}
\end{figure}

\begin{figure}
\includegraphics[width=\columnwidth]{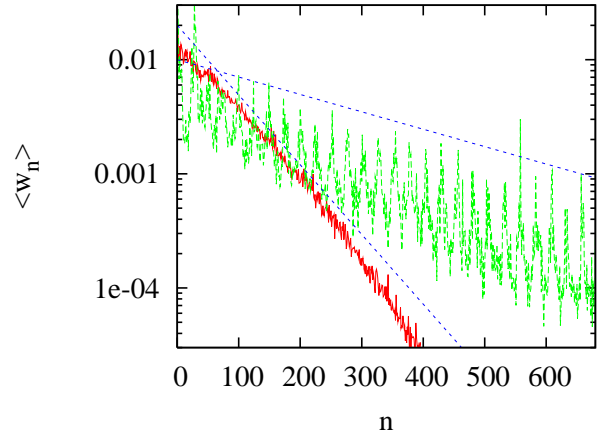}
\vglue -0.3cm
\caption{\label{fig2} 
(color online) Probability distribution $\langle w_n \rangle$ 
over the billiard eigenstates $n$
after a pulse of 2000 microwave field periods
(averaged over last 500 periods,
$n$ is counted from the Fermi level $n_F$).
Here $\epsilon = 1200$, $\omega=20$ (red/black curve)
and  $\epsilon = 300$, $\omega=160$ (green/gray curve).
Initially all probability is in the state $n=0$
shown in Fig.~1. The straight dotted lines show the
theoretical exponential decay with the localization length
(\ref{Eq3}).
}
\end{figure}
To check the validity of the above estimates and 
obtain all numerical coefficients we study numerically 
the dynamical photonic localization in the billiard
which is given by a conformal quadratic map of a unit circle 
proposed by Robnik \cite{robnik}.
Its shape in polar coordinates is $r(\phi)=R (1 + 2\lambda \cos\phi)$ 
with a parameter $\lambda \in [0,1/2]$. For $\lambda>1/4$ 
the billiard is non-convex
and almost fully chaotic, while for $\lambda=1/2$ it is rigorously known
to be ergodic and belongs to the class of K-systems \cite{sinai}. 
We restrict our studies mainly to the case with $\lambda=3/8$
shown in Fig.~1. For the quantum evolution we 
consider only even states of the billiard.
These states are symmetric with respect to reflection transformation 
$y$ to $-y$ and a microwave field linearly polarized along $x$-axis
gives transitions only inside this symmetry class.
We take the Fermi level to be $n_F=2001$ 
in this symmetry class (total quantum number around 4000) that
corresponds to $E_F=v_F^2/4=12586.2$ where for numerical simulations we
use the usual billiard units  with
$\hbar = R = 2 m = 1$. Then near $E_F$
the average level spacing for this symmetry class is $\Delta=6.266$.
The numerical method introduced in \cite{robnik}
allows to find efficiently the billiard eigenstates 
and the dipole matrix elements between them
(the eigenstate at the Fermi level is shown in Fig.~1).
Therefore, the quantum evolution induced by a microwave field
can be numerically integrated directly in the eigenbasis. 
This approach allows to follow  the
quantum excitation over thousands of microwave periods.
The integration time step was set to $\Delta t=10^{-4}$ 
and we ensured that its modification did not affect the excitation 
probabilities. In the numerical simulations the 
transitions below $n_F$ were suppressed, since in a dot all states 
below $n_F$ are occupied by electrons, and 
up to $n_{tot}=2000$ levels above $n_F$ were taken into account.
The escape in outcome lead is modeled as absorption of
all probability above  the level number $n_F+n_I$ after a microwave pulse
of finite duration. This corresponds to the tunneling barrier hight
$U=n_I \Delta$ between the dot and outcome lead.

Examples of probability distribution $w_n$ over billiard eigenstates
$n$ after a long microwave pulse are shown in Fig.~2. They
clearly show exponential localization of probability
which can be approximately described by $w_n \sim \exp(-2n/l)/l$.
The localization length in the basis $n$ is related to the 
photonic length $l_\phi = l \Delta / \omega$. For large $\omega=160$
the distribution shows a chain of equidistant peaks
corresponding to one-photon transitions
with the distance between peaks $\delta n = \omega/\Delta$.
For $\omega \sim \Delta$ the peaks disappear and $w_n$
decays in a homogeneous way.
\begin{figure}
\includegraphics[width=\columnwidth]{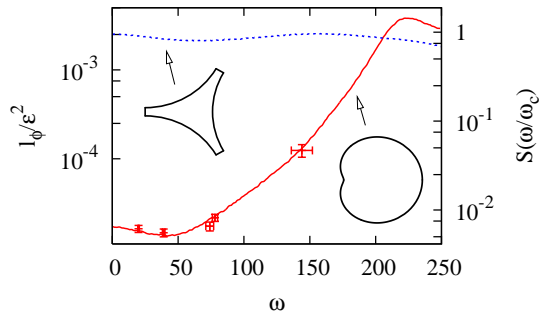}
\vglue -0.3cm
\caption{\label{fig3}
(color online) Dependence of the scaled localization length
$l_\phi/\epsilon^2$ (left axis) on the microwave frequency $\omega$. 
Numerical data 
for $\epsilon = 1200,1200,1200,600,300$ (from left to right) are 
shown by points with error bars, 
the full curve shows the theory
(\ref{Eq3}) with the classical spectral density $S(\omega/\omega_c)$
taken from \cite{prosen} for  the billiard of Fig.~1 (sketch),
here $\omega_c \approx 224$, 
$\Delta = 6.266$. Right axis shows the scale for $S(\omega/\omega_c)$.
The dashed curve shows the classical spectral density 
$S(\omega/\omega_c)$ for the 3-disks billiard (sketch);
$\omega$ is adjusted to the same collision frequency $\omega_c$
as for the full curve case.
}
\end{figure}

The localization length $l_\phi$ can be expressed through the
correlation function of dynamical motion inside the billiard.
Indeed, as discussed in
\cite{prosen} in the semiclassical regime the dipole matrix elements
of $x$ can be expressed via the spectral density of dynamical 
variable $x(t)$.
Using the definition of $D_E$ and the expression for
$\partial E/ \partial t$ we obtain
$D_E = \epsilon^2 \omega^2 R^2 S(\omega/\omega_c) /2\omega_c$.
Here $\omega_c=v_F/R$ and $S(\kappa)= 
\lim_{T\to \infty} \left|\int_{-T}^{T} d\tau
\xi(\tau) e^{{\rm i} \kappa \tau}\right|^2/2T$
is the dimensionless spectral density
of $\xi(\tau)=x(t)/R$ with $\tau = \omega_c t$ and $\kappa=\omega/\omega_c$.
Together with (\ref{Eq1}) this result leads to
\begin{equation}
\label{Eq3}
l_\phi =   \frac{\pi \epsilon^2 R^2}{\hbar \omega_c \Delta} 
S(\omega/\omega_c) \; .
\end{equation}
In the billiard units 
$\omega_c = 2\sqrt{E_F}$ and for our value of $E_F$
we have $\omega_c \approx 224$
and $l_\phi/\epsilon^2 \approx S/448$. The classical spectral density
$S(\omega/\omega_c)$ has been found numerically in 
\cite{prosen} and we show it in Fig.~3. As it is seen,
$S(\omega/\omega_c)$ has one maximum with $S(1) \approx  1.8$ 
and it goes to a constant value $S(0) \approx 0.004$ near zero.
Using (\ref{Eq3}) we can determine the quantum localization length
$l_\phi$ from the classical value of $S$. Without any fit
parameters this relation gives a good agreement with the numerical
data of Fig.~2. 

\begin{figure}
\includegraphics[width=\columnwidth]{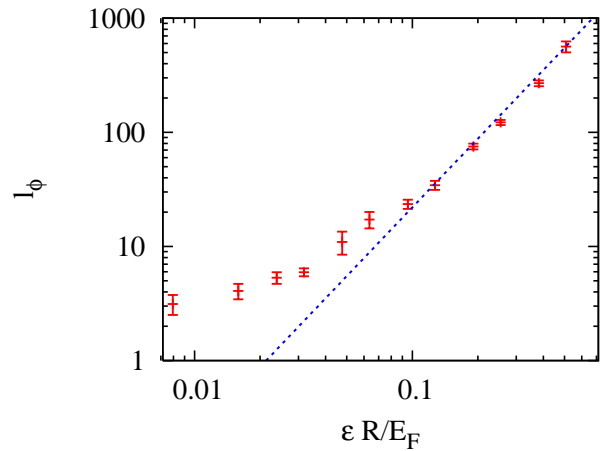}
\vglue -0.3cm
\caption{\label{fig4} 
(color online) Localization length $l_\phi$ 
as determined from numerical data 
(points) versus dimensionless
field strength $\epsilon R /E_F$ for  $\omega=20$.
The straight line gives the theoretical dependence (\ref{Eq3}).
}
\end{figure}

To check the theory (\ref{Eq3}) in more detail
we determine the value of $l_\phi$ from the fit of 
exponential probability decay:
$\langle w_n \rangle \propto \exp(-2 n \Delta/\omega l_\phi)$
($w_n$ is integrated in one-photon interval  to extract $l_\phi$).
The dependence of $l_\phi/\epsilon^2$ on $\omega$ is shown in Fig.~3
demonstrating a good agreement with the theoretical formula
(\ref{Eq3}) \cite{note}. The dependence on $\epsilon$ at a fixed
$\omega$ is shown in Fig.~4. It
is also in a good agreement with the theory (\ref{Eq3})
while $l_\phi \gg 1$ which is assumed by the semiclassical approach
\cite{note2}.

The nontrivial property of Eq.~(\ref{Eq3}) is that $l_\phi$
is directly dependent on the dynamical spectrum $S(\omega/\omega_c)$.
For the mesoscopic billiards discussed above $\omega \ll \omega_c$
and it is interesting to have billiards with large value $S(0)$.
Since $S(0)$ is simply the integrated auto-correlation function of
$\xi(\tau)$, a possible choice is a billiard formed by three touching
disks which exhibits a slow polynomial decay of correlations 
(see e.g. \cite{artuso}).
In Fig.~3  we show the classical power spectrum 
$S(\omega/\omega_c)$  versus $\omega$ for the 3-disks billiard 
(see sketch, the opening between touching circles has a length $0.05R$).
In this case the spectrum has a plateau at small $\omega$
that gives the value of $S(0)$ by a factor 250 larger
compared to the billiard of Fig.~1. The comparison of (\ref{Eq3})
with (\ref{Eq2}) gives $\chi = S(0)/4$ 
($a=2R$, $\nu_c=\pi \omega_c/2$). For the 3-disks billiard
we have $\chi \approx 0.25$ and $l_\phi \approx 1$ for 
$\epsilon = 0.5V/cm$ and ${\cal A}={\cal A}_0$, $n_e=n_{e0}$.

In fact the billiard shape is of primary importance for the efficiency
of microwave excitation. Indeed, a decrease of $\lambda$
from $\lambda=3/8$ to $\lambda=0.1$ makes the 
unperturbed dynamics inside
the billiard  quasi-integrable. In the latter case
a microwave field acts in an adiabatic way and the current through the
dot is reduced by orders of magnitude 
compared to the chaotic billiard as it is shown in Fig.~5
\cite{note1}.
This figure also shows that for  a chaotic billiard
the current through the dot
is strongly reduced as soon as 
$\hbar \omega l_\phi$
becomes smaller than the barrier hight $U=n_I \Delta$.
Therefore, a conductance $g$ between  leads
can be efficiently changed by varying a microwave field power.
The conductance appears due to diffusion in energy
which is rather similar to spatial diffusion in mesoscopic 
quasi-one-dimensional wires.
Using analogy with the latter case 
where $g=E_c/\Delta$ \cite{thouless}, 
we can write the microwave conductance of the billiard as
\begin{equation}
\label{Eq4}
g \sim l_\phi \exp (- 2 N_I/l_\phi)/N_I \; ,
\end{equation}
where $E_c = \hbar D_E/U^2$ is the Thouless energy for diffusion
in energy space, $N_I=U/\hbar \omega$ is the number of photons
required to pass over the lead barrier $U$
and an effective level spacing between quasienergy levels
is $\Delta_{ef} \sim  \Delta/N_I$.
As in \cite{thouless} it drops with the sample size $N_I$
and $g \sim E_c/\Delta_{ef}$. 
For $l_\phi \gg N_I$
the dot is in the metallic regime  while for $l_\phi \ll N_I$
it is insulating. Thus a microwave field 
can efficiently switch on and off
the conductance of a chaotic quantum dot blocked by a
tunneling barrier $U$
between the dot and outcome lead. 
To distinguish clearly the dependence (\ref{Eq4}) 
from the Arrhenius activation law 
at finite temperature $T$ we need to have $T \ll \hbar \omega l_\phi$.
\begin{figure}
\includegraphics[width=\columnwidth]{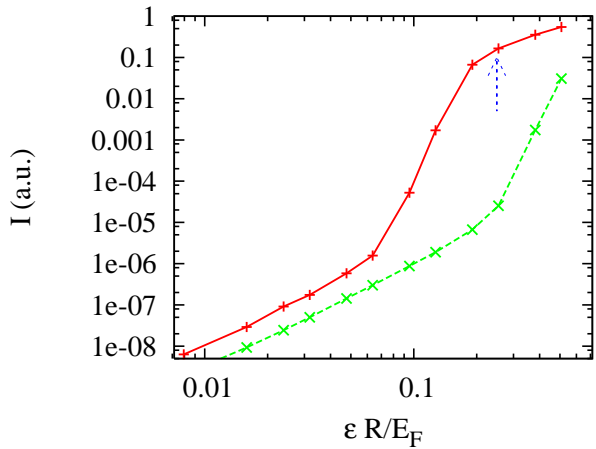}
\vglue -0.3cm
\caption{\label{fig5} Dependence of the current $I$ through 
a dot billiard on a scaled microwave field
strength $\epsilon R/E_F$ at $\omega =20$, $n_F=2001$. 
The current $I$, shown in arbitrary
units, is determined as the probability on 
billiard eigenstates, within
$2500 \leq n \leq 4000$, absorbed after a field pulse 
of 2000 microwave periods duration.
Two curves correspond to
fully chaotic billiard  at $\lambda=3/8$ (top)
and to  quasi-integrable billiard at $\lambda=0.1$ (bottom).
The arrow marks the position where the absorption border
is reached by localization
$n_I = 500 = \hbar \omega l_\phi/\Delta$.
}
\end{figure}

Of course, above we used a rather simplified model 
neglecting  interactions
between excited electrons and finite temperature
of a dot. These effects destroy quantum coherence
and  dynamical localization and  should be taken
into account when comparing the theory with the experiment.
However, we expect that a transition from metallic to
insulating behavior induced by a microwave field
(see (\ref{Eq4})) is a robust phenomenon and thus it can be
observed experimentally as it has happen with a microwave
ionization of Rydberg atoms.  
Present experimental techniques allow to 
observe  effects of microwave radiation 
on electron transport in mesoscopic dots (see e.g. \cite{kvon,bouchiat})
that makes possible  experimental investigations
of the effects discussed here.   

We thank Kvon Ze Don for discussions which originated this work.


\begin{thebibliography}{99}
\bibitem{note} Eq. (\ref{Eq3}) assumes that the energy excitation
        $\hbar \omega l_\phi$ remains small compared to $E_F$,
        at large $\omega$ it gives a noticeable fraction of $E_F$
        that leads to horizontal error bars in Fig.~3.
\bibitem{note2} Also the size of field oscillations should be larger than
       the wavelength implying $\epsilon/(m \omega^2) > R/n_F^{1/2}$.
\bibitem{note1} While in experiments it may be not easy to vary directly
     a dot shape, it may be possible to introduce a magnetic
     field which should make the dynamics quasi-integrable 
     as soon as the Larmor radius becomes much smaller than
     the dot size. For a dot with parameters ${\cal A}_0$ and $n_{e0}$
     discussed above this corresponds to a magnetic field $B \sim 1 T$.
\end{thebibliography}
\end{document}